\documentclass[trackchanges]{aastex6}
\usepackage{hyperref}
\usepackage{todonotes}
\usepackage{natbib}
\usepackage{amsmath}
\usepackage{comment}
\usepackage{multirow}
\begin{document}
\title{Constraints on Galactic Neutrino Emission with Seven Years of IceCube Data}
\author{
IceCube Collaboration:
M.~G.~Aartsen\altaffilmark{1},
M.~Ackermann\altaffilmark{2},
J.~Adams\altaffilmark{3},
J.~A.~Aguilar\altaffilmark{4},
M.~Ahlers\altaffilmark{5},
M.~Ahrens\altaffilmark{6},
I.~Al~Samarai\altaffilmark{7},
D.~Altmann\altaffilmark{8},
K.~Andeen\altaffilmark{9},
T.~Anderson\altaffilmark{10}
% ...
}
\altaffiltext{1}{Department of Physics, University of Adelaide, Adelaide, 5005, Australia}
\altaffiltext{2}{DESY, D-15735 Zeuthen, Germany}
\altaffiltext{3}{Dept.~of Physics and Astronomy, University of Canterbury, Private Bag 4800, Christchurch, New Zealand}
\altaffiltext{4}{Universit\'e Libre de Bruxelles, Science Faculty CP230, B-1050 Brussels, Belgium}
\altaffiltext{5}{Niels Bohr Institute, University of Copenhagen, DK-2100 Copenhagen, Denmark}
\altaffiltext{6}{Oskar Klein Centre and Dept.~of Physics, Stockholm University, SE-10691 Stockholm, Sweden}
\altaffiltext{7}{D\'epartement de physique nucl\'eaire et corpusculaire, Universit\'e de Gen\`eve, CH-1211 Gen\`eve, Switzerland}
\altaffiltext{8}{Erlangen Centre for Astroparticle Physics, Friedrich-Alexander-Universit\"at Erlangen-N\"urnberg, D-91058 Erlangen, Germany}
\altaffiltext{9}{Department of Physics, Marquette University, Milwaukee, WI, 53201, USA}
\altaffiltext{10}{Dept.~of Physics, Pennsylvania State University, University Park, PA 16802, USA}
% ...

\begin{abstract}
The origins of high-energy astrophysical neutrinos remain a mystery despite extensive searches for their sources.  We present constraints from seven years of IceCube Neutrino Observatory muon data on the neutrino flux coming from the Galactic plane.  This flux is expected from cosmic-ray interactions with the interstellar medium or near localized sources.  
Two methods were developed to test for a spatially-extended flux from the entire plane, both maximum likelihood fits but with different signal and background modeling techniques.  We consider three templates for Galactic neutrino emission based primarily on gamma-ray observations and models that cover a wide range of possibilities.  
Based on these templates and an unbroken $E^{-2.5}$ power-law energy spectrum, we set 90\% confidence level upper limits constraining the possible Galactic contribution to the diffuse neutrino flux to be relatively small, less than 14\% of the flux reported in \citet{Aartsen:2015knd} above 1 TeV.
A stacking method is also used to test catalogs of known high energy Galactic gamma-ray sources.
\end{abstract}
\section{Introduction} \label{sec:intro}
The high-energy sky is dominated by diffuse photon emission from our Galaxy, the first discovered steady source of astrophysical gamma rays~\citep{1968ApJ...153L.203C}. 
Cosmic-ray interactions with ambient interstellar gas are the dominant production mechanism for high energy gamma rays in the plane of the Galaxy via the decay of neutral pions.  
Diffuse neutrinos from the plane of the Galaxy are expected from these same interactions via decay of charged pions.
We perform searches for diffuse neutrino emission based on models constructed from gamma-ray observations.  

Substantial contributions near the Galactic plane are possible from discrete Galactic sources.  These sources can appear as point-like or could have noticeable spatial extensions, as for nearby supernova remnants (SNR).  We focus on catalogs of SNR and pulsar wind nebulae (PWN), all observed by gamma-ray observatories which are sensitive above 1~TeV.  A stacking analysis is performed on the sub-categories of these catalogs, taking into account the spatial extension and relative source strength where information is available.  

The IceCube in-ice array~\citep{2017JInst..12P3012A} is a Cherenkov detector that consists of 5160 digital optical modules (DOMs) deployed in the glacial ice under the South Pole between depths of 1.45 and 2.45~km.  Each DOM contains a 10'' photomultiplier tube \citep{2010NIMPA.618..139A} and associated electronics~\citep{2009NIMPA.601..294A}.  These DOMs are frozen into the ice along 86 vertical strings, each with 60 DOMs.  Of these strings, 78 have $\sim$125~m spacing on a triangular grid.  The remainder make up the denser DeepCore region.  

The IceCube collaboration has reported the detection of a flux of high energy ($>$10~TeV) astrophysical neutrinos~\citep{IceCubeCollaboration:2013hx,PhysRevLett.113.101101,Aartsen:2015knd,Aartsen2:2015rwa,DiffusePaper}.
In a combined fit to all available IceCube data, the flux was characterized from 25~TeV to 2.8~PeV as a power law with spectral index 2.50$\pm$0.09 \citep{Aartsen:2015knd}.  A recent analysis of only muon neutrinos in the northern sky, with a higher energy threshold of 191~TeV and sensitive up to 8.3~PeV, yields a harder spectral index of 2.13$\pm$0.13~\citep{DiffusePaper}.  This difference could indicate either a spectral break or a spatial anisotropy.  Both would be consistent with a relatively soft Galactic contribution dominating in the southern sky in addition to a harder, isotropic extragalactic component to the flux. 

The astrophysical neutrino signal is so far compatible with isotropy despite a large number of searches that have been performed trying to identify its origins~\citep[e.g.][]{Aartsen:2016oji}.  
Several extragalactic candidates have been shown to have a sub-dominant contribution to the flux.  
Notably, blazars are constrained to contribute less than 27\% of the flux for $E^{-2.5}$ or 50\% if the spectrum is as hard as $E^{-2.2}$~\citep{Aartsen:2016lir}.
Prompt emission from triggered gamma-ray bursts are strongly constrained to $<$1\% contribution to the astrophysical flux~\citep{2017arXiv170206868A}. 
In the case of starburst or star-forming galaxies, only a small percentage are cataloged.  The evidence in this case is indirect, but in order to avoid having the parent cosmic-ray population overproduce the {\it Fermi}-LAT extragalactic gamma-ray background, the contribution of star-forming galaxies to the diffuse neutrino flux must be sub-dominant~\citep[e.g.][]{Bechtol:2015wn}.  

There are some indications of an association of neutrinos with the Galactic plane.  In the three-year sample of IceCube high-energy starting events (HESE), some correlation with the Galactic plane was observed with a chance probability of 2.8\% ~\citep{PhysRevLett.113.101101}.
\citet{Neronov:2015uz}, using the public HESE data\footnote{\url{https://icecube.wisc.edu/science/data}}, explore the addition of an energy cut to the HESE sample and find a $>$3$\sigma$ correlation with the Galactic plane.
They found an optimum energy threshold of 100 TeV, where events have a higher probability of having an astrophysical as opposed to an atmospheric origin.  

The ANTARES neutrino detector, located in the Mediterranean Sea, has good sensitivity for the Galactic center region.  They perform a search for muon neutrinos in the region of the Galactic ridge.  For the case of a neutrino flux that extends into the GeV energy range as an $E^{-2.5}$ power law, they set a per flavor flux normalization (at 100~GeV) upper limit of $1.9 \times 10^{-17}$ GeV$^{-1}$ cm$^{-2}$ s$^{-1}$ sr$^{-1}$ in the region Galactic longitude $l<|40^{\circ}|$ and Galactic latitude $b<|3^{\circ}|$ (encompassing 0.145 sr).  This excludes the possibility of three or more of the HESE events coming from the central part of the Galactic plane~\citep{Adrian-Martinez:2016fei}.


In order to search for a diffuse Galactic signal of astrophysical neutrinos from the Galactic plane we employ two different analysis methods using partly overlapping data sets. Both methods use muon neutrinos and are primarily sensitive to the outer galaxy in the Northern hemisphere.
The first method is an extension of the standard point-source search method \citep{Aartsen:2016oji}. It is an unbinned maximum likelihood method that uses a template of the Galactic plane for a signal expectation and scrambled experimental data for the background estimation.  The method uses data from the full sky but is primarily sensitive to the Northern hemisphere. In the following it is referred to as the {\it ps-template} method. 
The second method is an extension of the method of the diffuse astrophysical neutrino measurement \citep{DiffusePaper}. It is based on binned multi-dimensional templates of all contributing flux components from atmospheric and astrophysical neutrinos. This method inherently includes systematic uncertainties but requires higher purity with respect to atmospheric muon background and is thus restricted to the Northern hemisphere. The standard two-dimensional templates in \citet{DiffusePaper} are based energy and declination. They have been extended to include the right ascension as the third dimension. In the following this method is referred to as the {\it diffuse-template} method.

The paper is organized as follows.
Section~\ref{sec:models} describes models for gamma-ray and neutrino emission from the Galaxy.  
Section~\ref{sec:method} gives details on the statistical methods used.
The constraints on the Galactic flux from the spatial template and stacking searches are given in Section~\ref{sec:results}, and our conclusions are given in Section~\ref{sec:conclusions}. 

\section{Models of Galactic Neutrino Emission} \label{sec:models}

\subsection{Diffuse Emission Models} \label{sub:diffuse_models}

Models for diffuse gamma-ray production in the Milky Way have steadily improved over time in order to keep pace with gamma-ray instruments, such as the {\it Fermi}-LAT~\citep{Ackermann:2012kna}.  These instruments have been able to generate high precision data sets that the models must reproduce~\citep[for a recent summary, see][]{Acero:2016qlg}.  Diffuse gamma-rays can be produced in electromagnetic processes such as inverse Compton and bremsstrahlung.  Alternatively, cosmic-ray interactions with the interstellar medium (ISM) can produce neutral pions, which decay to gamma rays.  A roughly equal number of charged pions is expected, leading to neutrino production in this case.  
Our analysis uses three spatial models that span a robust range of possibilities for Galactic neutrino emission: the {\it Fermi}-LAT $\pi^0$-decay template~\citep{Collaboration:2012vi}, the KRA-$\gamma$ (50 PeV cutoff) model~\citep{Gaggero:2015ce}, and a smooth parameterization of the Galaxy from \citet{Ingelman:1996md}.

The {\it Fermi}-LAT $\pi^0$-decay template is taken from the reference Galactic model in \citet{Collaboration:2012vi}.
There, the spectrum and composition of cosmic rays throughout the Galaxy is modeled assuming that cosmic rays propagate diffusively from a distribution of sources and are reaccelerated in the interstellar medium.  Model parameters are constrained by local observations of cosmic rays.  The targets for gamma-ray production are the interstellar radiation field and the interstellar gas.  The radiation field is modeled in two dimensions, the distance from the Galactic center and the height above the plane.  A fit to the {\it Fermi}-LAT gamma-ray data is used to determine the normalization of the interstellar radiation field intensity, which has considerable uncertainties.  The interstellar gas distribution is derived from radio measurements of the CO and HI line intensities.  Based on the measured radial velocity of the gas, the total gas mass is distributed over several concentric rings around the Galactic center.  The proportionality constant that relates the CO line emission to the molecular hydrogen gas density ($X_{\mathrm{CO}}$) is a free parameter in each of these rings, obtained in a fit to the gamma rays observed by the {\it Fermi}-LAT.  The total expected gamma-ray intensity in each direction is then obtained by integrating the gamma-ray yields from cosmic-ray interactions with the target over the corresponding line of sight. 
We extract just the $\pi^0$-decay component of this model and use it as a template for neutrino emission, shown in Figure~\ref{FIG:Method}. We do not convert the gamma-ray flux into an absolute prediction of the number of neutrinos but only consider the shape of the model.  The {\it Fermi}-LAT energy range where the model is validated is substantially lower than the energies of the neutrinos to which we are sensitive.  A hardening in the spectrum of the cosmic rays could substantially increase the neutrino predictions.  

\citet{Gaggero:2015ce} noticed that the model above does, in fact, under-predict the amount of gamma rays above a few GeV in the Galaxy, especially for higher-energy observations of the H.E.S.S and Milagro collaborations. 
They investigate ways to explain this residual flux by relaxing the constraint that cosmic-ray propagation is uniform in the Galaxy.  By allowing for a diffusion coefficient that depends on Galactic radius and an advective wind, they construct the KRA-$\gamma$ model, which matches the anomalous gamma-ray data better.  Others have also explored the potential for the {\it Fermi}-LAT $\pi^0$-decay signal to serve as a means to measure the cosmic-ray spectrum throughout the Galaxy and found similar evidence for spectral hardening, be it towards the Galactic center~\citep{Acero:2016qlg} or in the entire plane~\citep{Neronov:2015wx}.

Though the KRA-$\gamma$ model uses an independent cosmic-ray propagation code, it is based on the same underlying model of the ISM as the {\it Fermi}-LAT $\pi^0$-decay template.  This makes the spatial features very similar between the two models, but the KRA-$\gamma$ predictions are higher on average and more concentrated in the Galactic center region.  The most optimistic model, KRA-$\gamma$ with 50~PeV cosmic-ray cutoff, predicts 213 neutrino events in our 7-year sample.  

The most noticeable difference with respect to the $\pi^0$-decay template is in the part of the Galactic plane closest to the center but visible from the northern sky.  This is the region ($30^{\circ}<l<65^{\circ}$ and $-2^{\circ}<b<2^{\circ}$) where Milagro is sensitive and used to tune the model.
Note that the ARGO-YBJ~\citep{ARGOYBJ:1999aa} experiment reports that gamma rays between $\sim$350~GeV and $\sim$2~TeV are consistent with the {\it Fermi}-LAT model after masking out all sources~\citet{2015ApJ...806...20B}.
If the Milagro diffuse flux measurement above $\sim$1 TeV in the plane can be resolved into individual leptonic sources by HAWC, the KRA-$\gamma$ predictions may need to be adjusted downwards.  

Finally, we also consider a smooth parameterization of the Galaxy from \citet{Ingelman:1996md}.  This model lacks the detailed cosmic-ray modeling and mapping of the ISM of the first two models but captures the overall shape and structure of the Galaxy.  
The scale height of the Galaxy is also higher than the {\it Fermi}-LAT or KRA-$\gamma$ models.  
Though this model is cruder than the others, we view it as valuable to include since it gives us robust results that only depend on simple assumptions.

The model assumes that pure-proton cosmic rays are uniformly distributed throughout the entire volume and that the normalization and spectrum match that observed at Earth.  
The ISM extends out to a radius of 12 kpc with a density of $1.0~e^{-h/(0.26~\mathrm{kpc})}$ nucleon/cm$^3$
, where $h$ is the height out of the Galactic plane in either direction.  From simulations, we find that the model predicts 248 neutrino events in the 7-year muon sample.  Even with the higher number of events than the other models, these neutrinos follow an $E^{-2.7}$ spectrum up to the cosmic-ray knee and softening to $E^{-3.0}$ above that, which is closer to the atmospheric background spectrum than the other models.  Note that this prediction now takes neutrino oscillations, which the model predates, into account.

\subsection{Catalogs for Stacking} \label{sub:stacking_models}
Five different Galactic catalogs, each containing 4 -- 10 sources described in Table~\ref{TAB:Stacked Sources}, were examined with the standard point-source stacking technique seen in \citet{Abbasi:2010rd}.  The sources were grouped into smaller catalogs under the assumption that the sources within each category would have similar properties such as spectral index. This is important as the spectral index, although a free parameter in the analysis, is assumed to be the same for all sources in each catalog. These catalogs are based on results from Milagro~\citep{Atkins:2004jf}, HAWC~\citep{Abeysekara:2013tka}, and those compiled by SNR Cat~\citep{SNRcat:2012}.

The HAWC catalog consists of 10 sources observed by HAWC after collecting the first year of data in the inner Galactic plane \citet{Abeysekara:2015qba}. This catalog was inspired by the large overlap in sensitive energy range between the HAWC and IceCube detectors, permitting a multimessenger view of the same candidate astrophysical particle accelerators.  

The Milagro catalog contains six Milagro sources in the Cygnus region originally reported by \citet{abdo2007tev} and modeled in \citet{Kappes:2009zza} and \citet{GonzalezGarcia:2009jc} as possible PeVatron candidates. This catalog has been used for stacking analysis previously using four years of IceCube data where a 2\% p-value was found \citep{Aartsen:2014cva}.  This analysis updates these results by adding an additional three years of data. 

The final three catalogs are sub-catalogs of a group of SNRs taken from SNR Cat which have been observed in the TeV region with an age less than 3000~years. The selection of young SNRs was inspired by results demonstrating that SNRs less than 3000~years old are more efficient accelerators in the TeV region~\citep{Naurois:ICRC2015}. This group was then divided into three subgroups of sources based on their observed environment: those with known molecular clouds, those with associated PWN, and those with neither.

\begin{table}[h]
\begin{center}
  \caption{\label{TAB:Stacked Sources} Source information for the five stacked catalogs.}
  \begin{tabular}{ r|c|c|c|c|c}
    Catalog & Associated Names & R.A ($^{\circ}$) & Declination ($^{\circ}$) & Extension $\sigma$ ($^{\circ}$) & Age (yrs)\\
    \hline
    \multirow{6}{*}{Milagro Six\tablenotemark{a}} 
     & MGRO J1852+01 & 283.12 & 0.51 & 0.0 & - \\
     & MGRO J1908+06 & 286.68 & 6.03 & 1.3 & - \\
     & MGRO J2019+37 & 304.68 & 36.70 & 0.64 & - \\
     & MGRO J2032+37 & 307.75 & 36.52 & 0.0 & - \\
     & MGRO J2031+41 & 307.93 & 40.67 & 1.5 & - \\
     & MGRO J2043+36 & 310.98 & 36.3 & 1.0 & - \\
     \hline
    \multirow{9}{*}{HAWC\tablenotemark{b}} 
    & HWC J1825$-$133 & 276.3 & $-$13.3 & 0.5 & - \\
    & HWC J1836$-$090c & 278.9 & $-$9.0 & 0.5 & - \\
    & HWC J1836$-$074c & 279.1 & $-$7.4 & 0.5 & - \\
    & HWC J1838$-$060 & 279.6 & 6.0 & 0.5 & - \\
    & HWC J1842$-$046c & 280.5 & $-$4.6 & 0.5 & - \\
    & HWC J1844$-$031c & 281.0 & $-$3.1 & 0.5 & - \\
    & HWC J1849$-$017c & 282.3 & $-$1.7 & 0.5 & - \\
    & HWC J1857+023 & 284.3 & 2.3 & 0.5 & - \\
    & HWC J1904+080c & 286.1 & 4.44 & 0.5 & - \\
    & HWC J1907+062c & 286.8 & 6.2 & 0.5 & - \\ 
     \hline
    \multirow{10}{*}{SNR with mol. cloud\tablenotemark{c}} 
     & Tycho & 6.33 & 64.15 & 0 & 443\\
     & IC443 & 94.3 & 22.6& 0.16 & 3000\\
     & SN 1006 SW & 225.7 & $-$41.9 & 1.06 & 1009\\
     & HESS J1708$-$410 & 258.4 & $-$39.8 & 1.36 & 1000\\
     & HESS J1718$-$385 & 259.5& $-$37.4& 0.15 & 1800\\
     & Galactic Centre Ridge & 266.4& $-$29.0 & 0.2& 1200\\
     & HESS J1813$-$178 & 274.5 & $-$15.5& 0.77 & 2500\\
     & HESS J1843$-$033 & 281.6 & $-$3.0& 0 & 900\\
     & SNR G054.1+00.3 & 292.6 & 18.9& 0 & 2500\\
     & Cassiopeia A & 350.9 & 58.8& 0 & 316\\
     \hline
    \multirow{9}{*}{SNR with PWN\tablenotemark{c}} 
     & Crab& 83.6& 22.01& 0& 961\\
     & RX J0852.0$-$4622& 133.0& $-$46.3& 0.7& 2400\\
     & MSH 15$-$52 & 228.6& $-$59.1& 0.11& 1900\\
     & HESS J1634$-$472& 249.0& $-$47.3& 0.63& 1500\\
     & HESS J1640$-$465 & 250.3& $-$46.6& 0.87& 1000\\
     & SNR G000.9+00.1 & 266.8& $-$28.2&0 & 1900\\
     & HESS J1808$-$204 & 272.9& $-$19.4& 0.14& 960\\
     & HESS J1809$-$193 & 273.4& $-$17.8& 0.92& 1200\\
     & HESS J1825$-$137 & 278.4& $-$10.6& 1.63& 720\\
     \hline
    \multirow{4}{*}{SNR alone\tablenotemark{c}} 
     & RCW 86& 220.8& $-$62.5& 0.98& 2000\\
     & HESS J1641$-$463& 250.3& $-$46.3& 0.62& 1000\\
     & RX J1713.7$-$3946& 258.5& $-$38.2& 0.65& 350\\
     & HESS J1858+020& 284.5& 2.2& 0.08& 2300\\
    \hline
  \end{tabular}
  \tablenotetext{a}{\citet{Kappes:2009zza}}
  \tablenotetext{b}{\citet{Abeysekara:2015qba}}
  \tablenotetext{c}{\citet{SNRcat:2012}}
  \tablecomments{The stacking method uses a Gaussian distribution to represent either the source shape or the localization uncertainty.
  }
\end{center}
\end{table}

\section{Analysis Methods} \label{sec:method}

Maximum likelihood techniques are widely used in neutrino astronomy, and all analyses presented here use variations on existing methods.  The stacking method used to search the catalogs in Section~\ref{sub:stacking_models} is fully described in \citet{Abbasi:2010rd}.  The spatial template methods are newer and described below.  

\subsection{PS-Template Analysis} \label{subsec:unbinnedmethod}
The ps-template analysis method is a modification of the unbinned maximum likelihood analysis commonly employed in IceCube collaboration point-source searches~\citep{2008APh....29..299B,Aartsen:2016oji}. 
The analysis uses an event-wise point-spread function (PSF).  While the angular resolution of our best reconstructed events is small ($\sim$$0.1^{\circ}$) compared to the spatial structures of the Galactic plane, accounting for the PSF of our less well reconstructed events ($\sim$$3.0^{\circ}$) is important.  
The first modification is to account for the extension of the source by mapping the changing detector acceptance and convolving the true source hypothesis with the PSF of the events (in contrast to the delta-function source hypothesis used in point-source searches). 
The other modification relates to the estimate of the background using data.  In a point-source analysis, a hypothetical source has a very small contribution in the declination band and is treated as negligible for determining the background.  For the Galactic plane, the signal may extend over the entire sky and is no longer negligible.  We construct a {\it signal-subtracted} likelihood that acknowledges this contribution, making a small correction to the method introduced in \citet{Aartsen:2015xej}.  

As in \citet{Aartsen:2016oji}, the mixture model likelihood is defined as
\begin{equation} \label{eq:1}
L(n_s,\gamma) = \prod_{i=1}^{N}\left( \frac{n_s}{N} S_i(\mathbf{x}_i, \sigma_i, E_i; \gamma) + (1-\frac{n_s}{N})B_i(\sin \delta_i, E_i) \right),
\end{equation}
where $n_s$ is the number of signal events for a flux following spectral index $\gamma$; $N$ is the total number of events in the sample; $S_i(\mathbf{x}_i, \sigma_i, E_i; \gamma)$ is the signal probability distribution function (PDF) for event $i$ at equatorial coordinates $\mathbf{x}_i = (\alpha_i,\delta_i)$ with Gaussian PSF of width $\sigma_i$ and energy proxy $E_i$; and $B_i$ is the background PDF.  
In this case, the background PDF does not come directly from the observed data, which are now treated as a mixture of signal and background:
\begin{equation}
\tilde{D}_i(\sin \delta_i, E_i) = \frac{n_s}{N} \tilde{S}_i(\sin \delta_i, E_i) + (1-\frac{n_s}{N}) B_i(\sin \delta_i, E_i).
\end{equation}
The $\tilde{D}$ and $\tilde{S}$ terms are constructed by integrating the events in a small declination bin ($\sim$$1^{\circ}$) over right ascension to determine the event density as a function of $\sin \delta$ and $E$ for the experimental data and simulated signal, respectively.  Solving for $B_i$ and substituting into equation~\ref{eq:1} this gives the final signal-subtracted likelihood function as
\begin{equation}
L(n_s,\gamma) = \prod_{i=1}^{N}\left( \frac{n_s}{N} S_i(\mathbf{x}_i, \sigma_i, E_i; \gamma) + \tilde{D}_i(\sin \delta_i, E_i)  - \frac{n_s}{N} \tilde{S}_i(\sin \delta_i, E_i) \right).
\end{equation}
Though the signal and background PDFs are defined event-wise, in practice events in the same stable data taking periods are grouped together to construct the signal and background PDFs.  

\begin{figure}[htb]
\begin{tabular}{cc}
\includegraphics[width=.45\textwidth, keepaspectratio]{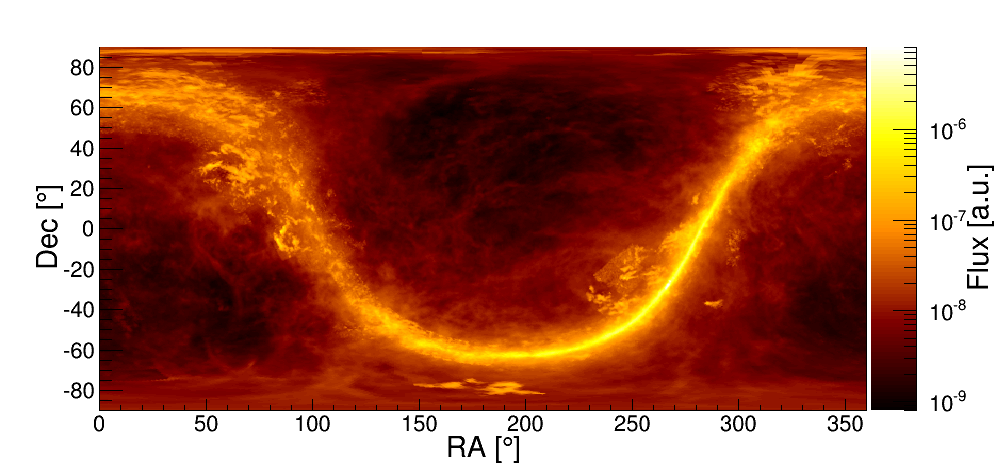} & \includegraphics[width=.45\textwidth, keepaspectratio]{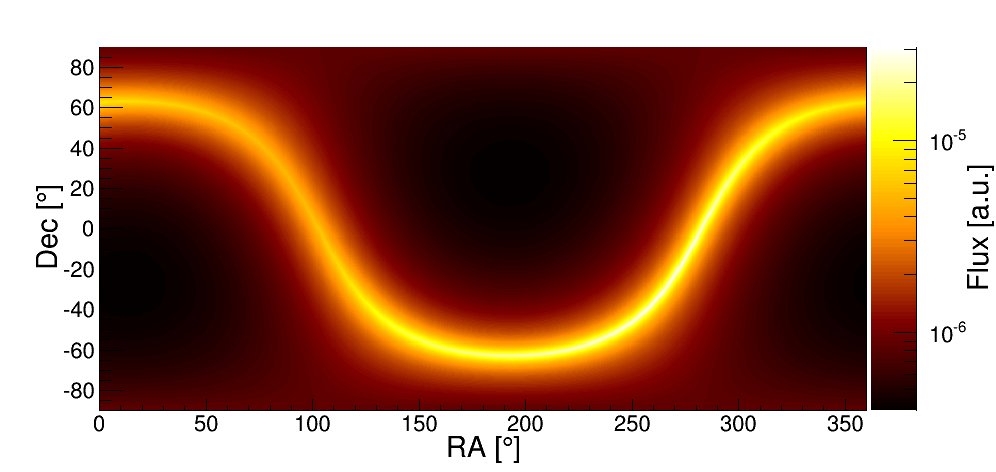} \\
(a) KRA-$\gamma$ (50 PeV cutoff) template & (b) Ingelman \& Thunman template\\
\includegraphics[width=.45\textwidth, keepaspectratio]{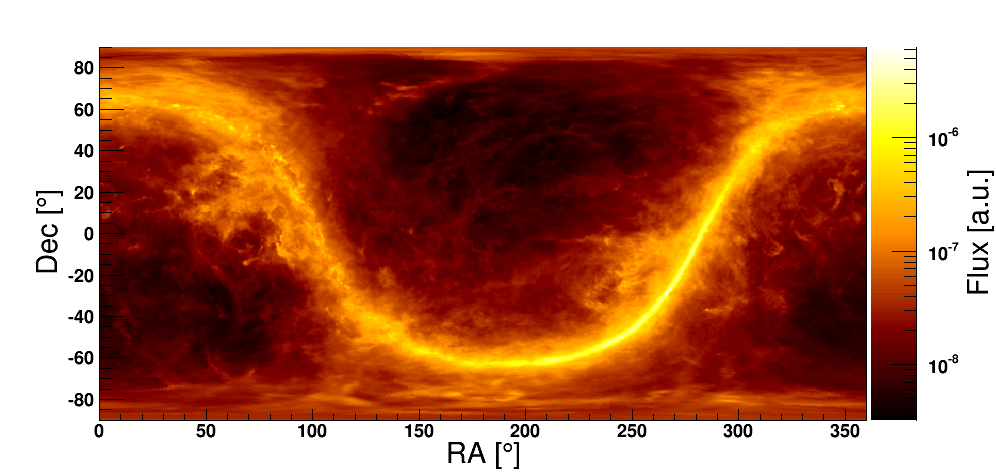} & \includegraphics[width=.45\textwidth, keepaspectratio]{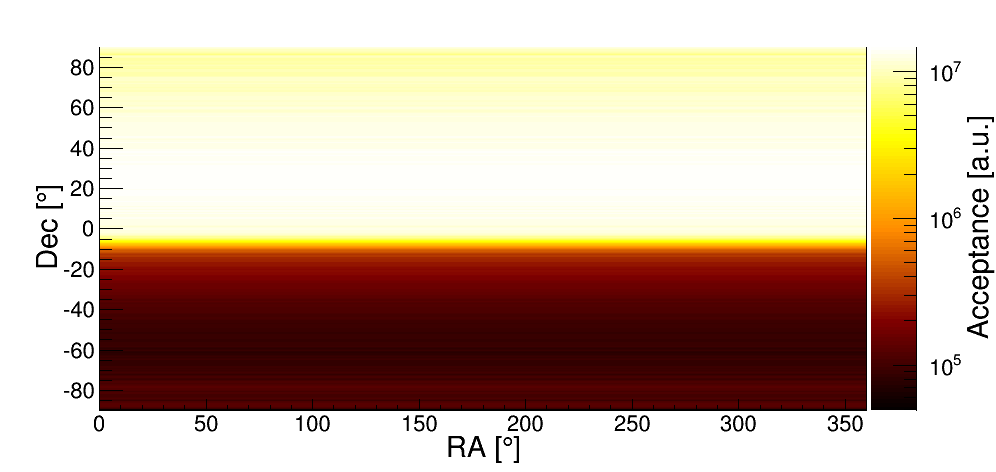} \\
(c) {\it Fermi}-LAT $\pi^0$-decay template & (d) Detector acceptance to $E^{-2.5}$ signal\\
\includegraphics[width=.45\textwidth, keepaspectratio]{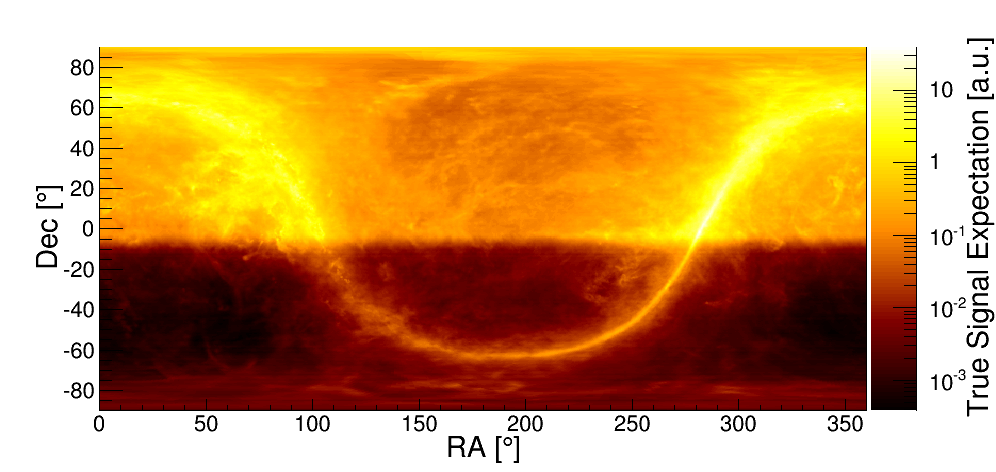} & \includegraphics[width=.45\textwidth, keepaspectratio]{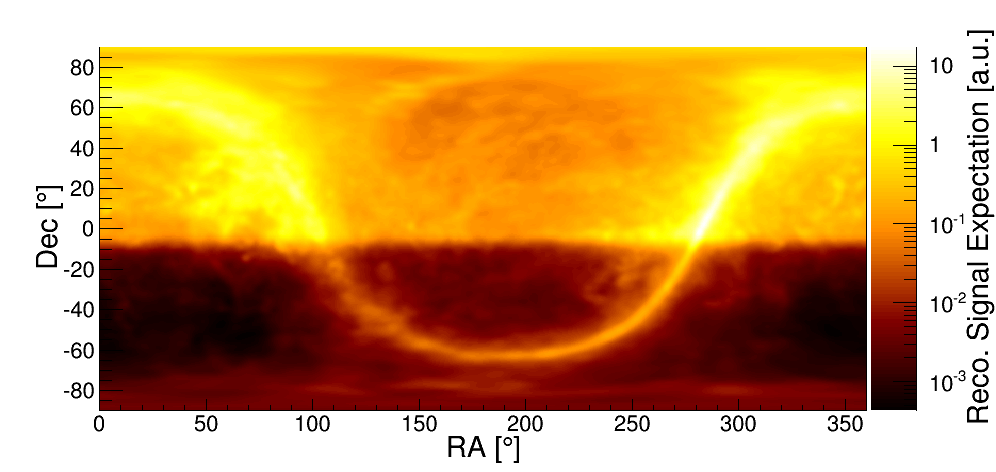} \\
(e) Signal PDF in true coordinates & (f) Signal PDF in reconstructed coordinates for $1.0^{\circ}$ PSF\\
\end{tabular}
\caption{\label{FIG:Method}
Three models for diffuse Galactic neutrino production described in Section~\ref{sec:models} are shown in equatorial coordinates: (a) KRA-$\gamma$ (50~PeV cutoff), (b) Ingelman \& Thunman parameterization, and (c) {\it Fermi}-LAT $\pi^0$-decay map. The models are followed by an illustration of the ps-template search technique described in Section~\ref{subsec:unbinnedmethod}: (d) the detector acceptance for an $E^{-2.5}$ power law, (e) the distribution of neutrino events before and (f) after smearing due to the median PSF with the $\pi^0$-decay map.  }
\end{figure}

The $S_i$ terms, which encode information about both the raw signal expectation and the detector performance, are constructed as follows.  Starting with a model for how the neutrino flux is distributed across the sky, we perform a bin-by-bin multiplication with the effective area to obtain the expected number of neutrinos per unit solid angle in the sample as a function of the direction and energy.  The effective area is determined using detailed simulations, described in \citet{DiffusePaper}.  
Then the map is convolved with the PSF of the event, which is adequately described by a Gaussian distribution of width $\sigma_i$ estimated for each event and ranging from 0.1$^{\circ}$ to 3.0$^{\circ}$.  In practice, maps are convolved in steps of 0.1$^{\circ}$ over this range.  
These steps are illustrated in Figure~\ref{FIG:Method}, integrating over energy for the case of a Galactic flux following an $E^{-2.5}$ power law.

A nested log-likelihood ratio between the best-fit signal strength and the null-hypothesis (no Galactic signal) is used to construct the test statistic.  
Under the null hypothesis and in the large sample limit the test statistic follows a half-$\chi^2$-distribution as expected~\citep{Cowan:2010js}.  
Final upper limits, sensitivities (given as median upper limits) and significances, such as the p-value of the experimental test statistic, are always calculated from scrambled data.
We use a strict 90\% upper limit construction~\citep{Neyman1967}.

The data sample, used by the ps-template as well as the stacking searches, is described in~\citet{Aartsen:2016oji}. It is an all-sky sample that spans seven years with a total live time of 2431 days and 730130 events.  Some of the data have been collected during the construction phase of IceCube with the partially completed detector.  The sensitivity comes primarily from the northern sky, where IceCube sees a wide energy range of neutrino-induced muons.  Though the sample extends into the southern sky, 
the sensitivity here is limited to very high energies since energy selection and veto techniques are used that reject the softer background of down-going muons from cosmic-ray air showers.  
For the benchmark case of an $E^{-2.5}$ spectrum, the energy range that contains 90\% of signal events is 400 GeV to 170 TeV and the median PSF is 0.79$^{\circ}$. 

\subsection{Diffuse-Template Analysis} \label{subsec:binnedmethod}

The binned maximum likelihood analysis is an extension of the analysis presented in \citet{DiffusePaper}. There, the contributions from conventional atmospheric neutrinos~\citep{Honda}, prompt atmospheric~\citep{ERS} and isotropic astrophysical neutrinos, assuming a power-law energy spectrum, are fitted to experimental data. The events are binned according to the reconstructed zenith angle and an energy proxy. The resulting histograms are analyzed using a maximum likelihood approach.
Each bin is modeled by a Poissonian likelihood function: 
\begin{equation}
L_i(\boldsymbol{\theta}, \boldsymbol{\xi}) = e^{-\mu_i(\boldsymbol{\theta}, \boldsymbol{\xi})} \cdot \frac{\mu^k_i}{k!},
\end{equation}
where $\boldsymbol{\theta} = \boldsymbol{\theta}(\Phi_{\mathrm{astro}}, \gamma_{\mathrm{astro}}, ...)$ describe the signal parameters (i.e. properties of the astrophysical fluxes) and $\boldsymbol{\xi}$ describe the nuisance parameters.
The expected number of events in bin $i$, $\mu_i$, is given by the sum of the four flux-expectations for the conventional, isotropic astrophysical, prompt, and Galactic flux:
\begin{equation}
\mu_i(\boldsymbol{\theta}, \boldsymbol{\xi}) = \mu^{\mathrm{conv}}_i (\xi_{\mathrm{conv}}, \xi_{\mathrm{det}}) + \mu^{\mathrm{astro}}_i(\Phi_ {\mathrm{astro}}, \gamma_{\mathrm{astro}}, \xi_{\mathrm{det}}) + \mu^{\mathrm{prompt}}_i (\xi_{\mathrm{prompt}}, \xi_{\mathrm{det}}) +\mu^{\mathrm{Galactic}}_i(\Phi_{\mathrm{Galactic}}, \xi_{\mathrm{det}}),
\end{equation}
where $\xi_{\mathrm{conv}}$ and $\xi_{\mathrm{prompt}}$ refer to nuisance parameters taking into account the theoretical uncertainties on the respective fluxes and $\xi_{\mathrm{det}}$ refers to nuisance parameters taking into account detector uncertainties. For more information on those parameters we refer to \citet{DiffusePaper}.
The final, global likelihood is the product of all per-bin likelihoods $L = \prod_i L_i$.

Compared to \citet{DiffusePaper}, this analysis is extended by including the reconstructed right ascension, thus changing the histograms from two to three dimensions. Additionally, a template for the Galactic contribution, $\mu^ {\mathrm{Galactic}}$, is added to the fit. Note that for the {\it Fermi}-LAT $\pi^0$-decay template the expected Galactic neutrino flux also depends on the Galactic spectral index.

In contrast to the method described in the previous section, this method models the expected contributions of every flux component using Monte Carlo simulations. 
This allows us to see how the isotropic component changes with the best-fit Galactic component. 
The test statistic is defined as a log-likelihood ratio in the same fashion as the previous method, with the same limit and significance calculations.  


The data sample for the diffuse-template analysis is described in \citet{DiffusePaper}. Compared to the previous method, the sample has a significantly higher purity of $>$99.7\%  with comparable effective area and a slightly improved PSF.  The sample is, however, limited to the Northern hemisphere where the high neutrino purity standards can be achieved.  
The time period is somewhat shorter as this selection does not apply to the first year when IceCube had just 40 of the final 86 strings deployed.  
The data set spans six calendar years with a total live time of 2060 days and 354792 events.
For the benchmark case of an $E^{-2.5}$ spectrum, the energy range that contains 90\% of signal events is 420 GeV to 130 TeV and the median PSF is 0.69$^{\circ}$. 

\section{Results} \label{sec:results}


\subsection{Constraints on Diffuse Emission in the Plane} \label{subsec:diffuseresults}

\begin{figure}[htb]
\centering\includegraphics[width=.6\textwidth, keepaspectratio]{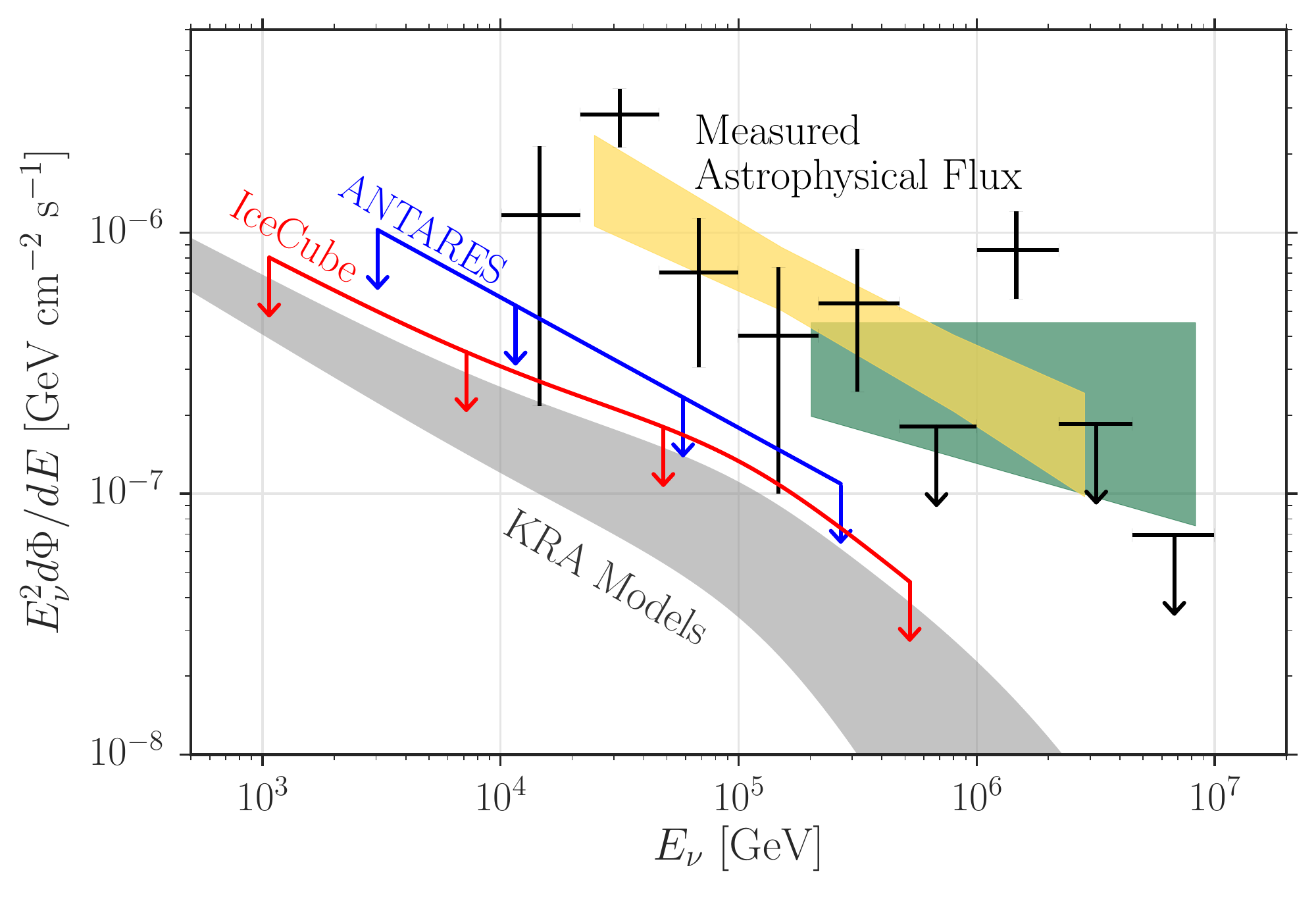}
\caption{\label{FIG:ULs} Upper limits at the 90\% confidence level on the three-flavor (1:1:1 flavor ratio assumption) neutrino flux from the Galaxy with respect to KRA model predictions and the measured astrophysical flux.  Fluxes are integrated over the whole sky for uniform comparison.  The IceCube 7-year upper limit for the KRA-$\gamma$ (50 PeV) model test for the ps-template method is shown in red.  The energy range of validity is from 1~TeV to 500~TeV.
This range is calculated by finding the low and high energy thresholds where removing simulated signal events outside these values decreases the sensitivity by 5\% each.
The ANTARES limit in blue~\citep{Adrian-Martinez:2016fei}, directly applicable in the Galactic center region ($-40^{\circ}<l<40^{\circ}$ and $-3^{\circ}<b<3^{\circ}$), has been scaled to represent an all-sky integrated flux for comparison. 
Due to ANTARES being located in the Northern hemisphere, it has a relatively high sensitivity in the southern sky near the Galactic center region using muon neutrinos.
The range of predictions for all the KRA models~\citep{Gaggero:2015ce} is shown as the gray band with the top of that band representing the KRA-$\gamma$ (50 PeV) model.  For comparison, measurements of the all-sky diffuse flux are shown: a differential unfolding (black points) and a power-law unfolding (yellow band) of combined IceCube data sets from \citet{Aartsen:2015knd} as well as a measurement based only on northern sky muon data (green band) from \citet{DiffusePaper}.}
\end{figure}

The sensitivities and results of spatial template analyses are summarized in Table~\ref{TAB:GPresults}.
Some excess from the Galaxy is observed in all cases, though it is not statistically significant.  
Because of the better sensitivity, the ps-template method was assigned, in advance of unblinding the data, to be the main result.  The diffuse-template method acts as a cross-check.  
The systematic uncertainty on the flux in the case of the ps-template analysis is estimated to be 11\% based on \citet{Aartsen:2016oji}.  For the diffuse-template, the systematic uncertainty is included directly in the method.  
The upper limit for the KRA-$\gamma$ test is shown in Figure~\ref{FIG:ULs} in comparison to the ANTARES upper limit, the KRA family of predictions, and the isotropic diffuse neutrino flux.  

\begin{table}[h]
\begin{center}
  \caption{\label{TAB:GPresults}Summary of results for both Galactic plane analysis methods for each of the three models.}
  \begin{tabular}{ c|c|c|c|c||c|c|c }
     & \multicolumn{4}{c||}{ps-template method} & \multicolumn{3}{c}{diffuse-template method} \\
    Spatial Template & $n_s$ & p-value & Sensitivity $\phi_{90\%}$ & Upper Limit $\phi_{90\%}$& p-value & Sensitivity $\phi_{90\%}$ & Upper Limit $\phi_{90\%}$\\
    \hline
    {\it Fermi}-LAT $\pi^0$-decay, $E^{-2.5}$ & 149 & 37\% & 2.97$\times$10$^{-18}$ & 3.83$\times$10$^{-18}$ & 7.0\% & 3.16$\times$10$^{-18}$& 6.13$\times$10$^{-18}$ \\
    KRA-$\gamma$ (50~PeV) & 98 & 29\% & 79\% & 120\% & 6.9\% & 95\% & 170\% \\
    Ingelman \& Thunman & 169 & 41\% & 220\% & 260\% & 19.8\% & 260\% & 360\% \\
    \hline
  \end{tabular}
  \tablecomments{Best-fit number of signal events $n_s$. 
  Fluxes are integrated over the full sky and parameterized as $d\phi_{\nu_{\mu}+\bar{\nu}_{\mu}}/dE = \phi_{90\%} \cdot (E/$100 TeV$)^{-2.5}$ GeV$^{-1}$ cm$^{-2}$ s$^{-1}$ with 90\% confidence level upper limits and median sensitivities quoted for $\phi_{90\%}$ or as a percentage relative to the model prediction. }
\end{center}
\end{table}

The Galactic excesses are somewhat larger and more significant for the diffuse-template cross-check, and this difference was investigated carefully for the benchmark $\pi^0$-decay template where the 7\% p-value was found.  
Part of the difference comes from the additional year of data used by the ps-template method.
If we restrict this method to the same time period, the
p-value drops from 37\% to 22\%.  
Running the ps-template method on the sample used by the diffuse-template method yields
a p-value of 21\%.  
Due to the purity requirement of the diffuse-template method, the check of the diffuse-template method on the alternative data set is not possible.

The spectrum of the signal neutrinos is given by the model in the cases of KRA-$\gamma$ and Ingelman \& Thunman, but we test a range of spectral hypotheses using the $\pi^0$-decay spatial template.  For a flux $\propto$$E^{-\gamma}$, the spectral index range that we test is quite broad, $1 < \gamma < 4$.  This is a wider range than we would expect in our standard models. However, it matches the range used in previous point source and stacking searches and allows for unexpected contributions, such as an unresolved population of hard or soft spectrum sources. The results of this coarse scan in spectral index are given in Table~\ref{TAB:GPresults2}.  
The small, best-fit Galactic component has a slight preference for $E^{-2.0}$ compared to $E^{-2.5}$, consistent with the results of the cross-check shown in the next section.

\begin{table}[h]
\begin{center}
  \caption{\label{TAB:GPresults2}Summary of results for the ps-template analysis scan in spectral index using the {\it Fermi}-LAT $\pi^0$-decay spatial template and power-law flux $\propto$$E^{-\gamma}$.}
  \begin{tabular}{ r|c|c|c|c}
    $\gamma$ & $n_s$ & p-value & Sensitivity $\phi_{90\%}$ & Upper Limit $\phi_{90\%}$\\
    \hline
    Softest: 4.0 & 13 & 45\% & 5.53$\times$10$^{-21}$ & 6.03$\times$10$^{-21}$ \\
    3.5 & 0 & 48\% & 7.27$\times$10$^{-20}$ & 7.27$\times$10$^{-20}$ \\
    3.0 & 15 & 49\% & 7.20$\times$10$^{-19}$ & 7.30$\times$10$^{-19}$ \\
    2.5 & 149 & 37\% & 2.97$\times$10$^{-18}$ & 3.83$\times$10$^{-18}$ \\
    2.0 & 103 & 24\% & 2.50$\times$10$^{-18}$ & 4.10$\times$10$^{-18}$ \\
    1.5 & 2 & 47\% & 3.79$\times$10$^{-19}$ & 3.79$\times$10$^{-19}$ \\
    Hardest: 1.0 & 0 & 76\% & 1.04$\times$10$^{-20}$ & 1.04$\times$10$^{-20}$ \\
    \hline
  \end{tabular}
  \tablecomments{Best-fit number of signal events $n_s$.    
  Fluxes are integrated over the full sky and parameterized as $d\phi_{\nu_{\mu}+\bar{\nu}_{\mu}}/dE = \phi_{90\%} \cdot (E/$100~TeV$)^{-\gamma}$ GeV$^{-1}$ cm$^{-2}$ s$^{-1}$ with 90\% confidence level upper limits and median sensitivities quoted for $\phi_{90\%}$. }
\end{center}
\end{table}

\subsection{2D Likelihood Scan and Implications for the Isotropic Astrophysical Flux} \label{subsec:diffuse_iso_impl}

\begin{figure}[htb]
 \centering\includegraphics[width=.70\textwidth, keepaspectratio]{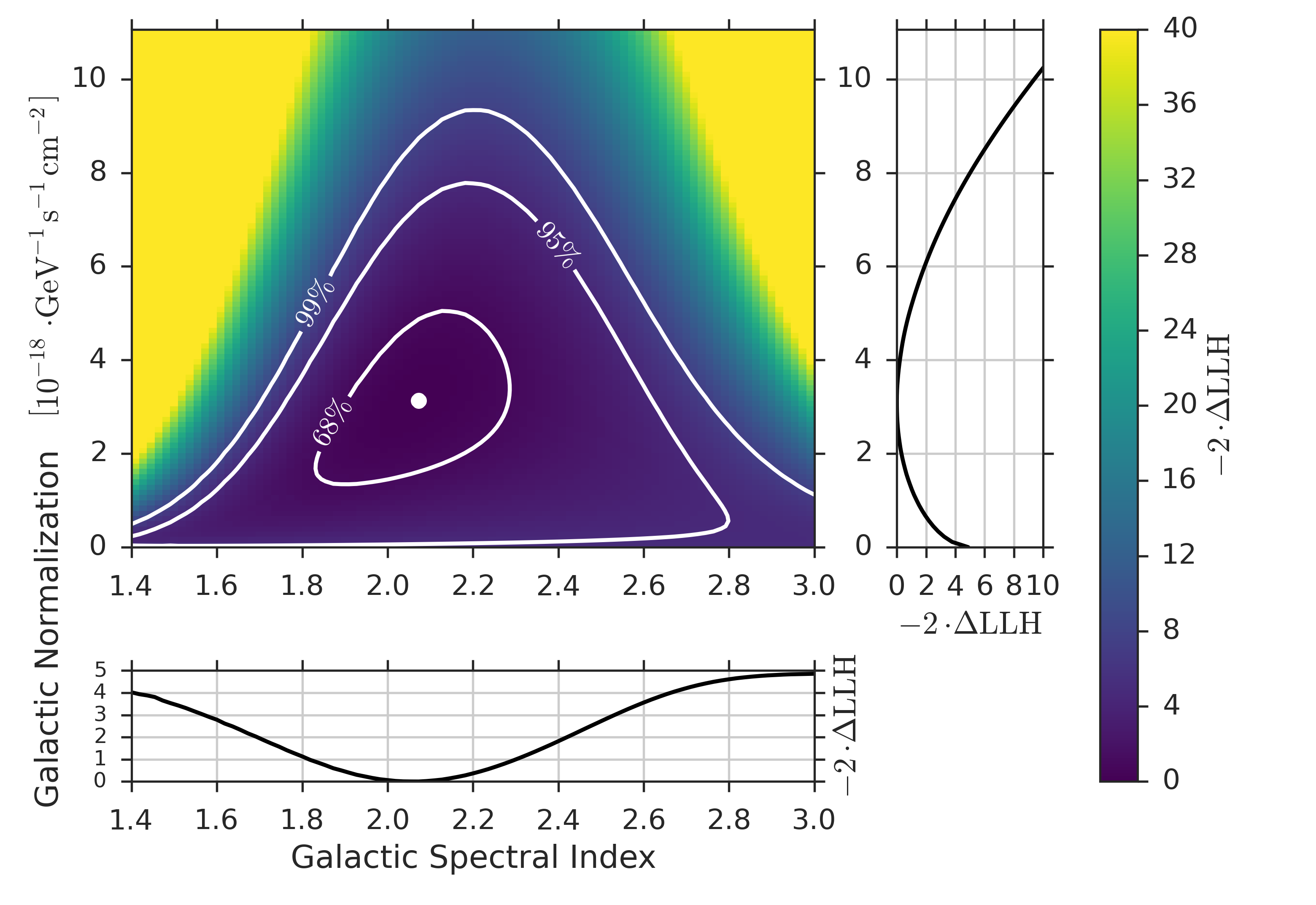}
  \caption{\label{FIG:2DScan_gal} 2D likelihood scan of spatially-integrated Galactic flux ($\nu_{\mu}+\bar{\nu}_{\mu}$) normalization and Galactic spectral index. The marginal curves show the respective one-dimensional profile likelihoods. A change in $-2\cdot \Delta$LLH of 1 corresponds to the 1$\sigma$ range.}
\end{figure}

Using the diffuse-template method a two-dimensional profile likelihood scan of the Galactic normalization and Galactic spectral index is performed for the benchmark {\it Fermi}-LAT $\pi^0$-decay template.  The results of this scan are shown in Figure~\ref{FIG:2DScan_gal}. The best-fit spectral index is $\gamma_{\mathrm{Galactic}} = 2.07 $, and the best-fit flux ($\nu_{\mu}+\bar{\nu}_{\mu})$ normalization at 100\,TeV is $\Phi_{\mathrm{Galactic}} = 3.13 \cdot 10^{-18}$GeV$^{-1}$cm$^{-2}$s$^{-1}$. The confidence contours have been estimated using Wilks' Theorem \citep{wilks1938}, whose applicability has been confirmed at several points in the parameter space with Monte Carlo pseudo-experiments.
The best-fit spectral index is found to be somewhat harder than expected from gamma-ray observations.  However it is consistent with an index of 2.5 (2.7) at the 1.6 (2.1) $\sigma$-level.

\begin{figure}[htb]
\centering\includegraphics[width=.70\textwidth, keepaspectratio]{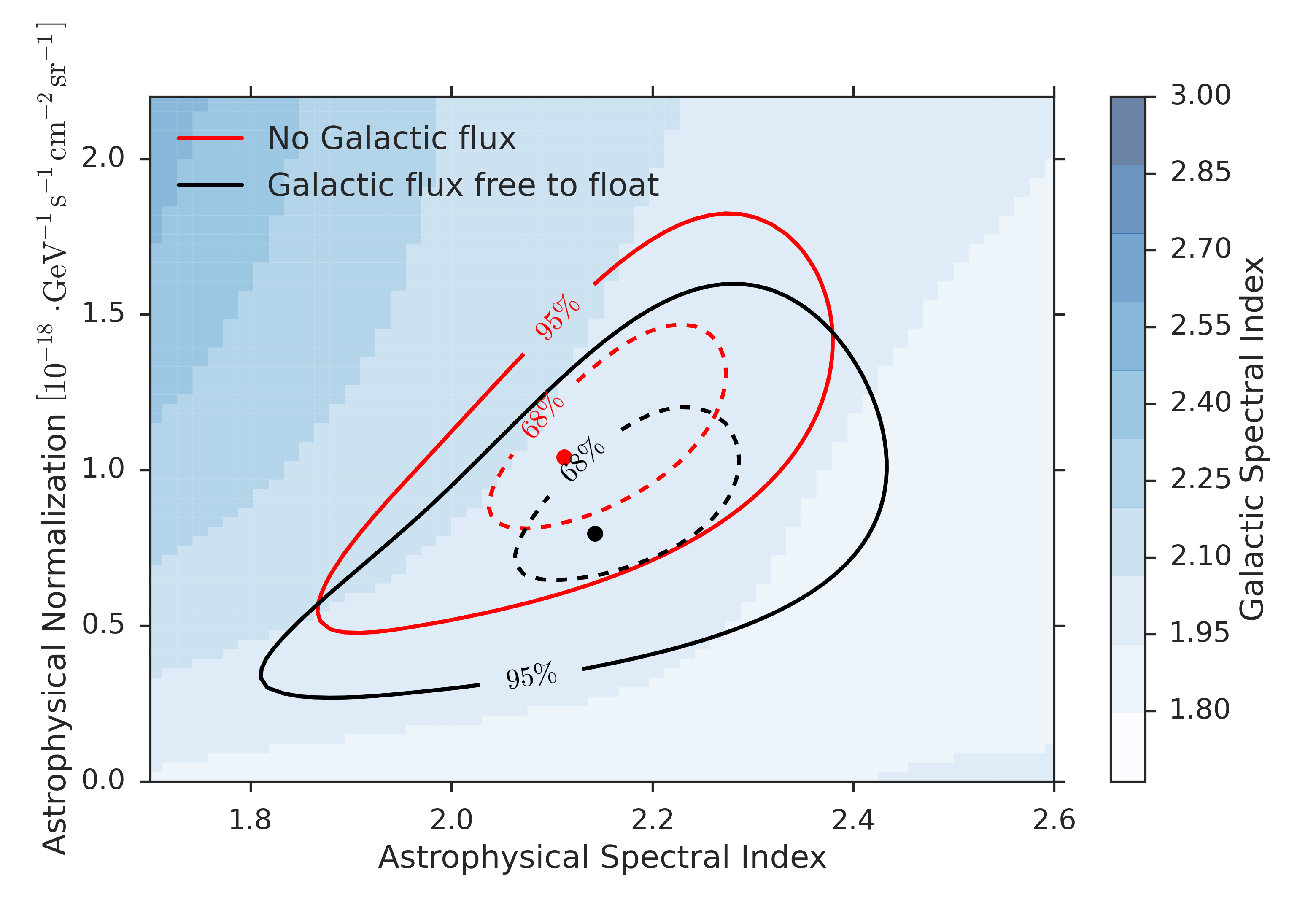}
\caption{\label{FIG:2DScan_astro_gamma_astro} Contours of isotropic astrophysical flux ($\nu_{\mu}+\bar{\nu}_{\mu}$) normalization and spectral index with and without Galactic flux in the fit.  The case with no Galactic flux differs a small amount from \citet{DiffusePaper} because we do not include $\nu_{\tau}$ here.}
\end{figure}

Since the diffuse-template method is an extension of the method used to characterize the isotropic astrophysical flux in \citep{DiffusePaper}, we can analyze the impact of allowing an additional Galactic component in the fit on the isotropic flux parameters. For this we perform a profile-likelihood scan of the isotropic flux normalization and spectral index while allowing the Galactic flux parameters to float freely at every scan point. Figure \ref{FIG:2DScan_astro_gamma_astro} shows the resulting likelihood contours in comparison to the contours obtained by restricting the Galactic flux to zero. The color scale shows the best-fit Galactic plane spectral index at each point in the scan. Although the additional freedom given by the Galactic flux nuisance parameter causes the isotropic normalization to decrease, the size of the contour grows only marginally. The hypothesis of zero isotropic flux is still excluded at $3.8\sigma$. This shows that the observation of an isotropic astrophysical signal is robust against a signal from the Galactic plane and that the latter can only contribute a sub-dominant fraction to the total observed extraterrestrial flux.



\subsection{Constraints on Source Catalogs} \label{subsec:stackingresults}

The results from the stacking analyses are shown in Table~\ref{TAB:StackingResults}.  All catalogs are consistent with small and statistically insignificant excesses.  The most significant result with a p-value of 25\% is the case of SNR with molecular clouds, which gives just 16.5 excess events and a very soft spectral index close to the limit of $\gamma<4.0$.  Compared to the results of~\citet{Aartsen:2015knd} the all-sky integrated upper limit assuming an $E^{-2.5}$ spectrum for all of these catalogs is found to be between 4 and 5 orders of magnitude below the fit for the isotropic diffuse flux.

The most promising Galactic catalog for stacked analysis was that of the six Milagro sources. The previous iteration of this search gave a p-value of 2\%~\citep{Aartsen:2014cva}.
The two sources MGRO~J1908+06 and MGRO~J2019+37 in this catalog are also two of the most significant of 74 individual source candidates investigated as part of the 7-year IceCube source list search~\citep{Aartsen:2016oji}.  Their respective pre-trial p-values are 0.025 and 0.23.  
However, the results of this search for the six Milagro sources showed a decrease in significance from a p-value of 2\% to 29\%. 
This result excludes the model of \citet{Kappes:2009zza}, based on Milagro observations, with more than $95\%$ confidence.  Given that MGRO~J1908+06 and MGRO~J2019+37 are two of the most significant results from the source list search in \citet{Aartsen:2016oji}, we investigated the apparent discrepancy.
In contrast to the source list search, this stacked analysis used source extensions on the order of 1$^{\circ}$, as reported by Milagro~\citep{abdo2007tev}.
The coordinates for this stacking analysis come from the Milagro data as opposed to those found by {\it Fermi}-LAT.  While the latter are better localized, the former come directly from $>$1~TeV gamma rays, which are a better match for the IceCube energy range.  The combination of these effects on the strongest source alone, MGRO~J1908+06, results in a substantial significance decrease, from a p-value of 4.6\% to 47\%.


\begin{table}[h]
\begin{center}
  \caption{\label{TAB:StackingResults} Summary of results for the Galactic catalogs. }
  \begin{tabular}{ c|c|c|c|c|c}
    Source Catalog & Number of sources & p-value & $n_s$ & $\gamma$ & Upper Limit $\phi_{90\%}$\\
    \hline
    Milagro Six & 6 & 30\% & 31.8 & 3.95 & 3.98$\times$10$^{-20}$\\
    HAWC Hotspots & 10 & 31\% & 17.3 & 2.38 & 9.48$\times$10$^{-21}$\\
    SNR with mol. clouds & 10 & 25\% & 16.5 & 3.95 & 2.23$\times$ 10$^{-19}$\\
    SNR with PWN & 9 & 34\% & 9.36 & 3.95 & 1.17$\times$10$^{-18}$\\
    SNR alone & 4 & 42\% & 3.82 & 2.25 & 2.06$\times$10$^{-19}$\\
    \hline
  \end{tabular}
  \tablecomments{$n_s$ and $\gamma$ are the best-fit number of signal events and spectral index, respectively.
  Fluxes are given as the sum over the catalog and parameterized as $d\phi_{\nu_{\mu}+\bar{\nu}_{\mu}}/dE = \phi_{90\%} \cdot (E/$100 TeV$)^{-2.5}$ GeV$^{-1}$ cm$^{-2}$ s$^{-1}$ with 90\% confidence level upper limits quoted for $\phi_{90\%}$.}
\end{center}
\end{table}

\section{Conclusions} \label{sec:conclusions}
We have presented searches for neutrino signals associated with the Galactic plane using seven years of IceCube muon neutrino data, focusing on diffuse emission from interactions of cosmic rays with the ISM.  
We are able to exclude that more than 14\% of the isotropic diffuse neutrino flux as measured in \citet{Aartsen:2015knd} comes from the Galactic plane for the case of the {\it Fermi}-LAT $\pi^0$-decay template and an $E^{-2.5}$ power law.
This assumes the flux continues down in energy to 1~TeV or less, as would be expected for the case of cosmic-ray interactions with the ISM.  The astrophysical neutrino flux has only been measured above 10~TeV so far, and its diffuse-template fit parameters are not changed significantly when the fit includes a Galactic component.  

Our measurement is primarily sensitive in the Northern hemisphere, where IceCube has a high efficiency for a wide energy range of muons induced by neutrinos.  Our limits are quoted assuming various all-sky spatial models of neutrino emission.  The KRA family of models span a wide range and imply that a Galactic neutrino contribution must be present at some level.
The most optimistic, KRA-$\gamma$ with a 50~PeV cosmic-ray cutoff, concentrates the most flux towards the Galactic center.  Even though a higher fraction of the flux is in the southern sky, our limits are just 20\% higher than this model prediction.  
In principle, if even more flux were concentrated near the Galactic center than in KRA-$\gamma$, it could be missed in this analysis and the limits violated.  However, it is difficult for this to happen without overproducing gamma rays~\citep{Kistler:2015oae,Gaggero:2015ce}.  One possibility to circumvent this is if production occurs very near Sgr~A* where the environment may be opaque to gamma rays~\citep{Kistler:2015oae}.  The ANTARES detector also sets
relevant constraints measured directly in the region surrounding the Galactic center ($-40^{\circ}<l<40^{\circ}$ and $-3^{\circ}<b<3^{\circ}$).  They limit the neutrino flux to be less than 60\% higher than the KRA-$\gamma$ model at 100~TeV~\citep{Adrian-Martinez:2016fei}.  

A blind search both for individual point sources and for multiple sub-threshold hotspots in the plane has previously been performed and sets constraints on the contribution of a small number of localized sources~\citep{Aartsen:2016oji}. 
The stacking analysis presented here improves these results for several catalogs where the source locations are known. The catalog results are all of low significance and allow us to exclude the model of \citet{Kappes:2009zza}.  Newer models motivated by more recent gamma-ray observations predict a lower flux with softer indices~\citep{Gonzalez-Garcia:2013iha}.

While our flux constraints focus on the plane of the Galaxy, there are still possibilities for the flux to originate in or very near the Galaxy.  The possibility of cosmic-ray interactions with a gas halo extending out to $\sim$100~kpc is still being actively explored~\citep{2013ApJ...763...21F,2014PhRvD..89j3003T,2016arXiv160807421K}.  Another possibility is the annihilation \citep{Aartsen:2015xej,2016arXiv161204595A} or decay \citep{Murase:2015tw} of dark matter particles in the Galactic halo.  For these hypotheses, the emission is much more isotropic than the Galactic emission templates that we tested.  

There are possibilities to improve the sensitivity for Galactic neutrino searches.  A search for point sources in the Southern hemisphere using cascade-like events in IceCube has been shown to have a sensitivity comparable to ANTARES~\citep{2017arXiv170502383I}.  A joint analysis that includes both the track and cascade channels will offer promising improvements.  
Additionally, a joint Galactic plane analysis between IceCube and ANTARES, similar to the joint point-source analysis that produced better limits in the southern sky~\citep{Collaboration:2015cp}, would provide the strongest constraints on neutrinos from the Milky Way with all available data.

\acknowledgments

We acknowledge the support from the following agencies:
U.S. National Science Foundation-Office of Polar Programs,
U.S. National Science Foundation-Physics Division,
University of Wisconsin Alumni Research Foundation,
the Grid Laboratory Of Wisconsin (GLOW) grid infrastructure at the University of Wisconsin - Madison, the Open Science Grid (OSG) grid infrastructure;
U.S. Department of Energy, and National Energy Research Scientific Computing Center,
the Louisiana Optical Network Initiative (LONI) grid computing resources;
Natural Sciences and Engineering Research Council of Canada,
WestGrid and Compute/Calcul Canada;
Swedish Research Council,
Swedish Polar Research Secretariat,
Swedish National Infrastructure for Computing (SNIC),
and Knut and Alice Wallenberg Foundation, Sweden;
German Ministry for Education and Research (BMBF),
Deutsche Forschungsgemeinschaft (DFG),
Helmholtz Alliance for Astroparticle Physics (HAP),
Initiative and Networking Fund of the Helmholtz Association,
Germany;
Fund for Scientific Research (FNRS-FWO),
FWO Odysseus programme,
Flanders Institute to encourage scientific and technological research in industry (IWT),
Belgian Federal Science Policy Office (Belspo);
Marsden Fund, New Zealand;
Australian Research Council;
Japan Society for Promotion of Science (JSPS);
the Swiss National Science Foundation (SNSF), Switzerland;
National Research Foundation of Korea (NRF);
Villum Fonden, Danish National Research Foundation (DNRF), Denmark

\bibliographystyle{apj}

\bibliography{references}
\end{document}